\documentclass[12pt,preprint]{aastex}

\newcommand{\kms}{km\,s$^{-1}$}
\newcommand{\sii}{[S~{\sc ii}]}
\newcommand{\Feii}{[Fe~{\sc ii}]}
\newcommand{\Ha}{H{$\alpha$}}

\newcommand{\twelveCO}{{}\mbox{$^{12}$}CO}

\newcommand{\thirteenCO}{{}\mbox{$^{13}$}CO}
\newcommand{\jone}{J=1$\rightarrow$0}

\newcommand{\jthree}{J=3$\rightarrow$2}
\newcommand{\vel}{\rm{v}}
\newcommand{\mv}{$M(\vel)$}
\begin{document}
\title{Entrainment Mechanisms for Outflows in the L1551 Star-Forming Region}
\shorttitle{Entrainment Mechanisms for Outflows in L1551}
\shortauthors{Stojimirovi\'c et al.}

\author{Irena Stojimirovi\'c, Gopal Narayanan, Ronald L. Snell} 
\affil{Astronomy Department, University of Massachusetts at Amherst,
  MA 01003}

\and

\author{John Bally}
\affil{University of Colorado, Boulder, CO}
\email{irena@nova.astro.umass.edu, gopal@astro.umass.edu, snell@astro.umass.edu}
\begin{abstract}

We present high sensitivity \twelveCO\ and \thirteenCO\ \jone\
molecular line maps covering the full extent of the parsec scale L1551
molecular outflow, including the redshifted east-west (EW) flow. We also 
present \twelveCO\ \jthree\ data that extends over a
good fraction of the area mapped in the \jone\ transition. We compare
the molecular data to widefield, narrow-band optical emission in
H$\alpha$. While there are multiple outflows in the L1551 cloud, the main
outflow is oriented at 50\arcdeg\ position angle and appears to be driven by
embedded source(s) in the central IRS~5 region. The blueshifted outflowing 
molecular gas extends to the edge of the molecular cloud and beyond the last 
HH object, HH 256. On the contrary, the redshifted molecular gas terminates 
within 
the cloud, short of the most distant HH object, HH 286, which lies well
beyond the cloud boundary. The \jthree\ data indicate that there 
may be molecular emission associated with the L1551 NE jet, within the 
redshifted lobe of main outflow. We have also better defined the previously 
known EW flow and believe we have identified its blueshifted counterpart. We 
further speculate that the origin of the EW outflow lies near HH 102. 
We use velocity dependent opacity correction to estimate the mass and the 
energy of the outflow. 
The resulting mass spectral indices from our analysis, are systematically
lower (less steep) than the power law indices obtained towards other
outflows in several recent studies that use a similar opacity
correction method. We show that systematic errors and biases in the
analysis procedures for deriving mass spectra could result in errors
in the determination of the power-law indices.  The mass spectral
indices, the morphological appearance of the position-velocity plots
and integrated intensity emission maps of the molecular data, compared
with the optical, suggest that jet-driven bow-shock entrainment is the
best explanation for the driving mechanism of outflows in L1551.
The kinetic energy of the outflows is found to
be comparable to the binding energy of the cloud and sufficient to 
maintain the turbulence in the L1551 cloud.
\end{abstract}


\section{Introduction}

From the emergence of a hydrostatic core, the collapse of a
protostellar core is accompanied by winds and mass loss
\citep{lada85} probably driven by magnetospheric accretion
\citep[e.g.][]{koenigl91,edwards94, hartmann94}.
Integrated over time, the effect of a wind from a young star is to
blow away the material that shrouds it during its earliest evolution. 
The discovery of bipolar
molecular outflows has been a key to the understanding of this process
\citep[e.g.][SLP hereafter]{slp80}. These molecular
outflows have dimensions of up to several parsecs, masses one or two
orders of magnitude larger than their driving sources, and tremendous
kinetic energies \citep{bally83}. Such massive flows must
represent swept-up material as the winds, emerging from the star
and/or its circumstellar disk, interact with their ambient medium.

One of the important unanswered questions in outflow studies
is the mechanism by which the jets/winds emerging from the
embedded star/disk entrain and accelerate the surrounding molecular
material. Currently there are three scenarios: the ``wide-angle wind''
model in which a wide-angled magnetized wind expands like a bubble and
sweeps up the ambient medium in a momentum conserving manner into the
shell at the wind-bubble ambient medium interface \citep{shu91, shu95,
shu00, ls96, mm99, lee00, lee01}. Then there are two types of ``jet''
entrainment models, the ''turbulent'' jet entrainment model, in which
ambient material is accelerated through a turbulent viscous mixing
layer that surrounds the jet formed by Kelvin-Helmholtz instabilities
\citep{canto91,stahler94, lizano95}; and the ''prompt'' entrainment
model in which bow shock formed in the jet head drives material ahead
and laterally to the axis of the flow \citep{deyoung86, raga93,
stone93a, stone93b, stone94, masson93, suttner97, zz97, ssy97, dr99,
lee01, lee02}.

It is essential to distinguish between these models to not only
understand the processes that govern the formation of molecular
outflows, but also the nature of the jet/wind launching mechanism and
the underlying accretion disk processes. Since it is difficult to
observe the jet/wind launching regions directly, the properties of
large scale bipolar outflows can shed some light on the nature of
their initial conditions. In addition, the energy and momentum
deposited by molecular outflows within the molecular cloud can be
large enough to power turbulence in the cloud or even disrupt the
parts of the cloud. Understanding outflow properties over the entire
outflow history is critical in determine their impact on the cloud.

Within the last decade, with the advent of CCD cameras with 20\arcmin\
fields-of-view, more than two dozen giant Herbig-Haro (HH) flows with
dimensions ranging from 1 to 7 pc have been discovered
\citep{rbd97}. HH flows are chains of optically visible shock-excited
nebulae which, demonstrate the existence, at least during certain
evolutionary phases, of exceedingly well collimated and highly
supersonic wind components \citep[e.g.][]{reipurth86}. The tremendous
size of parsec-scale outflows implies that molecular material is being
entrained at parsec-scale distances from the driving source, and thus
it may have a significant impact on the kinematics and energy density
of a substantial volume of the parent molecular cloud.

Our current estimates of the ''typical'' outflow properties may be far 
from accurate, since there is no large homogeneous sample of molecular 
outflows which have been mapped with good resolution and sensitivity 
\citep{richer2000}. Consequently, our current understanding of the 
processes might be heavily subjective, and biased by a few frequently 
studied examples. We have been involved in a multi-wavelength study of
 a dozen or so parsec-scale outflows, complementing narrow-band optical 
observations with wide-field, sensitive, multi-transitional \twelveCO\ 
and \thirteenCO\ observations (Stojimirovi\'c et al. 2006 in prep.). Here, 
we present the detailed results of our multi-wavelength study on the 
L1551 molecular outflow. 

The bipolar molecular outflow associated with the L1551 small dark
cloud in the Taurus molecular cloud complex, was the first recognized
bipolar outflow from a young stellar object (SLP) and is considered
one of the best examples of its kind. At a distance of 140 pc
\citep{lynds62, kdh94}, the L1551 dark cloud has been the site of many
multi-wavelength studies that reveal the dynamical complexity of this
low mass star formation region.

Optical images of the L1551 outflow region, show new distant HH object, 
HH 286 \citep{drb99}, making the whole extent of the optical 
emission in the L1551 flow 1.3 pc. We made sensitive, large 
spatial extent, high spatial and velocity resolution maps at 
millimeter and submillimeter wavelengths. Large-spatial extent mapping 
performed on parsec-scale outflows, allows us to study in detail the 
entrainment mechanisms by which jets/winds drive the observed 
molecular flows. With the availability of both \twelveCO\ and 
\thirteenCO\ data over the full spatial extent of the flow, we are able 
to take a {\em full} and proper accounting of the mass and hence 
energetics of the outflows in the L1551 system.  Comparing optical and 
millimeter data, we are able to discriminate among entrainment models 
based on morphological grounds.

\section{Observations}

\subsection{FCRAO CO \jone\ observations}

A full mapping of \twelveCO\ and \thirteenCO\ in the J = 1 $\rightarrow$ 0
transition has been performed in several observing seasons over a
three-year period with the SEQUOIA receiver at the Five College Radio
Astronomy Observatory (FCRAO) 14~m telescope. SEQUOIA \citep{ege99} is
a cryogenic focal plane array designed for the 85--115.6 GHz range,
and was upgraded in spring 2002 from the original 16 to 32 pixels. The
receiver is configured as a dual-polarized 4$\times$4 array. The
telescope's half-power beamwidths are 45\arcsec\ and 47\arcsec\ for
\twelveCO\ and \thirteenCO\ transitions respectively. The antenna 
temperatures were corrected for the main beam
efficiencies of 0.45 for \twelveCO\ and 0.5 for \thirteenCO. 

The initial observations were taken in 2001 using raster mapping mode
in only the \twelveCO\ \jone\ transition, with channel spacing of
0.21~\kms\ and velocity resolution of 0.25~\kms. The data were
baselined and convolved to the 22.5\arcsec\ Nyquist sample grid using
CLASS.  From 2002, observations were taken using On-The-Fly mapping
technique (OTF) and a dual channel correlator (DCC), which allows user
to observe 2 frequencies simultaneously with all 32 pixels at SEQUOA,
generating 64 spectra for each read of the DCC.  The orthogonal
polarizations of the SEQUOIA array are averaged to produce spectra
with better signal-to-noise ratio (S/N). The OTF data were reduced
with the ''OTFTOOL'' software. OTFTOOL is a suite of tools to inspect,
edit, regrid the raw otf data and transform the result into a CLASS or
FITS form, developed by M. H. Heyer, G. Narayanan and M. Brewer and
available for public usage. Using OTFTOOL, data were examined, channel
maps as well as individual spectra were checked for any scanning
artifacts, baselined and regridded to the 22.5\arcsec\ sampled
grid. RMS noise weighting was used to combine the data.  Regridding
takes all of the redundant measurements of the OTF map and
constructs the end product - convolved spectra on a regular grid
written into a CLASS file. For both \twelveCO\ \jone\ and \thirteenCO\
\jone\ transitions, the 50 MHz spectrometer bandwidth setting was used
resulting in 50 kHz channel spacing.

Within CLASS, OTF data were smoothed and resampled to the channel spacing
of the 2001 \twelveCO\ data. The datasets were combined and FITS format 
files were generated. The resulting velocity spacing is 0.21~\kms and 
pixel spacing is 22.5\arcsec.

The system temperatures (T$_{sys}$) in our observations range from
400 -- 700 K for \twelveCO\ and between 200 -- 500 K for
\thirteenCO. Regions mapped with the higher noise level were repeated,
combined and averaged in order to get a constant lower noise level
over the whole extent of the map, with resulting mean rms per 
velocity channel of 0.19~K for \twelveCO\ and 0.12~K for \thirteenCO. 
Antenna pointing and focus were checked every few hours and corrected 
using SiO masers.

The analysis were done both in the CLASS and IDL software of Research
Systems Inc. Detailed studies of the physical parameters
characterizing the outflows were performed using IDL.

\subsection{HHT CO \jthree\ observations}

Submillimeter \twelveCO\ and \thirteenCO\ observations in the \jthree\
transition were carried in December of 2001 and January of 2004 at the
10m Heinrich Hertz Telescope (HHT). HHT's half-power beamwidths are
22\arcsec\ and 23\arcsec\ for \twelveCO, \thirteenCO\ transitions
respectively. Main beam efficiency is 0.5 and forward scatter
efficiency of 0.9.  As a front-end, the SIS-345GHz receiver was used
in single channel mode to achieve the highest sensitivity in that
channel. We used the availability of the OTF mapping procedure and
mapped 5\arcmin$\times$5\arcmin\ maps with row spacing along
declination axis of 6\arcsec\ and 8\arcsec\ along RA (scanning rate
10\arcsec/sec). L1551 has been mapped in \twelveCO\ \jthree\ along the
major outflow axis.  Single \thirteenCO\ spectra towards the central
YSO and selected lines of sight were also taken. As a back-end
configuration, we used all three available AOS's: AOS A, AOS B and AOS
C, filter banks and Chirp spectrometer. The two 1 GHz bandwidth AOS's
have mean resolutions of 934 kHz for AOSA and 913 kHz for AOSB. The
250 MHz bandwidth of the AOSC has mean resolution of 385~kHz. The AOS's
use 2048 element linear CCD's and thus over sample the spectra by
about a factor 2 for AOSA and AOSB and a factor 3 for AOSC. The filter
bank spectrometer has 3 bandwidth modes: 256 channels of 1 MHz, 256
channels of 250 KHz, and 128 channels of 62.5 KHz resolution,
respectively.  The final data set is made using only AOSA and AOSB,
since AOSC and filter banks experienced noise problems. Spectral
resolution is 0.9~\kms. With T$_{sys}$ around 900 K, several repeats
resulted in mean rms of $\sim 0.1$~K for the data presented here.

\section{Morphology and Kinematics of Outflows}
\subsection{Overview of the L1551 Region}

Following the discovery of the Herbig-Haro objects, HH 28 and HH 29
\citep{herbig74}, and an embedded source, IRS 5 \citep{ssv76}, in the
L1551 dark molecular cloud \citep{lynds62}, SLP found a large bipolar
outflow centered on IRS 5. Multi-wavelength observations
of L1551 cloud conducted through the last 25 years show that the
region hosts several young stellar objects in different evolutionary
stages and multiple overlapping outflows. What was thought to be a
beautiful example of a single bipolar flow turns out to have at least
three outflows \citep{msw91, pb91, torrelles87}. Figure~\ref{fig:l1551_first} 
shows the blueshifted and redshifted
integrated intensity map of our \twelveCO\ \jone\ data overlayed on
the \Ha\ image, previously published by \citet{drb99}.

\subsubsection{Young Stellar Objects}

There are a number of star-formation sites within the L1551 cloud, some of 
the most studied are: IRS 5, L1551 NE and the HL Tau region.
IRS 5 is a deeply embedded class I source \citep{ssv76}, with bolometric 
luminosity of
38~$L_{\odot}$ \citep{emerson84}, hidden from the direct view by up to 150
magnitudes of visual extinction \citep{stocke88}. It is a close binary
system \citep{rodriguez86, rodriguez98, campbell88, bc85} with
components separated by $\sim$ 0.$^{\prime\prime}$35 \citep{lmw97}.
Located 149\arcsec\ east-northeast from the IRS 5 is L1551 NE
\citep{emerson84}, a Class 0 source with bolometric luminosity of
4.5~$L_{\odot}$. Radio continuum observations at 3.5 cm suggested
\citep{rar95} and later confirmed \citep{rrab02} that L1551 NE is
binary source with 0$^{\prime\prime}$.5 separation. Approximately
6\arcmin\ to the north from IRS 5, lies the HL Tau region, with two
well-studied T-Tauri stars, HL and XZ Tau. In addition HH 30* and 
LkH$\alpha$ 358 embedded stars are found in the same HL Tau region.

\subsubsection{Jets}

Numerous jets have been reported emerging from these three embedded
systems. IRS 5 which is believed to be the powering source for the
large bipolar molecular outflow has two clearly defined jets
\citep{fl98,itoh00,rodriguez03}. The measured position angles are
PA=247$^{\circ}\pm 3^{\circ}$ for the Northern IRS 5 jet and
PA=235$^{\circ}\pm 1^{\circ}$ for the Southern IRS 5 jet, using
3.5~cm VLA observations with resolution of 0.\arcsec1
\citep{rodriguez03}. Surprisingly, the position angle of
either jet powered by IRS 5 is not aligned with the PA=50\arcdeg\
morphology of the molecular CO outflow.  However, the Northern IRS 5
jet orientation points to two regions along the edges of the bipolar
flow where unusually bright and warm ``high-velocity'' CS emission has
been seen \citep{ps95,yokogawa03}. These regions are 4\arcmin\ west,
near HH 102, and 2\arcmin\ east from IRS 5, at L1551 NE position, and
have been hypothesized as working surfaces of the Northern IRS 5 jet.
It has been speculated that the formation of L1551 NE was induced by
the IRS 5 outflow impact \citep{emerson84, ps95, yokogawa03}.

The L1551 NE binary system is observed to drive a jet delineated
by \Feii\ emission 
\citep{reipurth00}. The bright \sii\ emission HH 454 knots lie within
the 2\arcdeg\ of the \Feii\ jet.  The HH 454 - \Feii\ jet axis has
PA $\sim$243\arcdeg, and we refer to this axis as L1551 NE
jet. Proper motion study of the HH 454 knots, show that blueshifted
knots are extending toward the southwest and redshifted knots are
found toward the northeast from L1551 NE \citep{drb99}. This means
that if L1551 NE is capable of entraining CO gas in the region, it
would be in the same general orientation as the IRS 5 outflow.

In the HL Tau region several optically identified jets of high
velocity gas are reported \citep{mbr88}. In the CCD images of
\citet{drb99}, at least four flows emerge at different angles from
four sources. HL Tau drives a jet at PA $\sim$ 50\arcdeg, while XZ Tau
drives a poorly collimated wind, traced by a set of expanding bubbles
imaged by HST towards PA $\sim$ 20\arcdeg\ \citep{krist99}. The HH 30*
source is located 1\arcmin.5 south of HL Tau and its optical jet,
stretches several arcmin from the source at PA $\sim$ 35/215\arcdeg.
LkH$\alpha$ 358 also appears to drive a flow, probably toward PA $\sim$
70/250\arcdeg\ and is a likely candidate for the origin of HH 265 object 
\citep{moriarty2006}.

\subsubsection{CO and Herbig-Haro Outflows}
  
Figure~\ref{fig:l1551_first} is
dominated by the well known bipolar structure which we will call the
main outflow.  The main outflow's blueshifted emission lies along PA
$\sim$230\arcdeg\, and extends 20\arcmin\ southwest from IRS 5. The
main outflow redshifted emission extends 16\arcmin\ north-east of IRS
5 along PA $\sim$50\arcdeg.  Weak redshifted emission is detected in
the far north-east corner of the map and may be related to a further
extension of the outflow.  In addition to the bipolar outflow
symmetric about IRS 5, there are a number of other features that
complicate the outflow.  First, blueshifted emission is detected near
the HL Tau region, well within the red lobe of L1551 flow, and
presumably related to outflow activity in the HL Tau  region \citep{msw91,
pb91, torrelles87}.  The most striking feature is the
previously discovered redshifted component oriented east-west (EW
flow) that lies north of the blueshifted main outflow \citep{msw91,
pb91}. Finally, a 
blueshifted feature, can be seen extending into the main redshifted 
outflow lobe to the north-east beyond HH 262 \citep{msw91} 
and may be connected to the EW flow; an idea that will be discussed in more 
detail later in the paper.

\citet{drb99} reported the discovery of HH 286, making the whole
extent of the optical flow 1.3 pc between HH 286 and HH 256. The
molecular outflow extends slightly beyond HH 256 in the south-west,
however the outflow terminates short of HH 286.  The full extent of
the CO outflow is approximately 32\arcmin\, which corresponds to 1.3 pc 
at the distance of L1551 cloud. Therefore the spatial
extents of the optical and CO outflows, are the same although displaced.  
Our CO maps extend well beyond HH 286 and in Figure~\ref{fig:12COcore} and
Figure~\ref{fig:13CO} we show the integrated intensity maps of the
\twelveCO\ and \thirteenCO\ \jone\ emission within the velocity range
of the line core.  The \thirteenCO\ \jone\ map of the line core emission 
provides our best measure of the distribution
of ambient velocity gas. Both figures clearly illustrate that there is no
molecular gas at the position of the HH 286.  Thus, the molecular
outflow ends short of HH 286 because there is no molecular material to
be entrained in the outflow.

\subsection{Velocity Structure of the Outflows \label{sec:region}}

In Figures~\ref{fig:l1551_first} and \ref{fig:CO32} we show the
integrated intensity maps of the redshifted and blueshifted outflowing
gas in the \twelveCO\ \jone\ and \jthree\ transitions overlayed on
an image of \Ha\ emission.  The spatial extent of the \jthree\
map is considerably smaller than that of the \jone\ transition.
Although the outflow appears more collimated in the
\jthree\ map, this is largely due to the restricted angular extent
of the map, which did not cover either the HL Tau or EW outflow regions.
The \jthree\ data is primarily useful in probing the
inner regions of the main flow at higher angular resolution.  

In the following three sections, we will discuss in detail the 
velocity field associated with the molecular gas within the different flows 
in L1551.  We will discuss the main, HL Tau, and EW outflows.
 
\subsubsection{Main Flow}

In Figures~\ref{fig:mosaic_blue}a and \ref{fig:mosaic_blue_submm}a 
the lowest velocity blueshifted emission is shown.  
In the larger \jone\ map (see
Figure~\ref{fig:mosaic_blue}a) one can identify the blue lobe
associated with the main outflow, a bow-shaped emission feature
associated with HL Tau, and a feature extending north-east from the
main flow toward HH 262.  The blue lobe of the main flow has a parabolic
shell structure that originates at IRS 5, with IRS 5 embedded within
it.  This parabolic shell encompasses most of the \Ha\ emission and is
best delineated in the higher angular resolution image of the \jthree\
transition of CO seen in Figure~\ref{fig:mosaic_blue_submm}. 

At higher blueshifted velocities the emission becomes more collimated
and is confined closer to the outflow axis.  The nested velocity
structure for the blue lobe of this outflow is very well known.  The
emission (see Figure~\ref{fig:mosaic_blue_submm}b,c) predominantly
arises in three clumps associated with HH 29, HH 259 and HH 102.  For
the clumps associated with HH 29 and HH 259, the strongest CO \jthree\
emission lags behind the optically defined shock regions.  At the
highest blueshifted velocities, (see Figures~\ref{fig:mosaic_blue}d and 
\ref{fig:mosaic_blue_submm}d), the emission is confined to a region
located toward the end of the outflow cavity defined by the optical
emission, near HH 259. 

The alignment of the HH 454a, HH 29, HH 259, HH 28 with L1551 NE jet axis 
and the proper motion studies of HH 28 and HH 29 have been used to suggest 
that they share a common origin in L1551 NE \citep{drb99}. 
If indeed this is the case, it is possible that some of the blue emission 
we see in the main flow may originate in material entrained by L1551 NE 
outflow.
    
In Figures~\ref{fig:mosaic_red} and \ref{fig:mosaic_red_submm},
we show mosaics of integrated intensity channel maps of the
redshifted \twelveCO\ emission. Besides
the main flow, at the lowest redshifted velocities in the larger
\jone\ map, the beginning of the EW flow can be seen.  The velocity
structure in the redshifted gas of the main flow shows some
similarities to that of the blue lobe.  Again, there is evidence for 
a nested velocity structure, however the exact interpretation is
complicated by the presence of HL Tau.  In
Figure~\ref{fig:mosaic_red}a one can see an extension of redshifted
emission from IRS 5 to the north towards HL Tau.  This emission could
either be part of a parabolic shell morphology of the main red lobe,
or emission associated with the HL Tau outflow.  
As in the blue lobe, the
higher velocity emission is offset from IRS 5 and confined closer to
the outflow axis and the highest velocity emission is located near the
end of the outflow near HH 262.

Maybe the most tantalizing evidence of molecular emission associated with 
L1551 NE jet is seen in Figure~\ref{fig:mosaic_red_submm}a,b where there is 
a linear emission feature that arises close to NE and extends for approximately
5\arcmin\ to the north-east, close to the HH 262.  The position angle of this 
linear feature is $\sim$ 60\arcdeg\, similar to the orientation of the \Feii\ 
jet driven by L1551 NE \citep{reipurth00}. A similar suggestion was made by 
\citet{moriarty2006}. However, the feature could instead 
be just part of the parabolic shell associated with the main flow from IRS 5.
The radial velocity of HH 262 indicates that it is redshifted 
\citep{lopez98} and that it may be either related to L1551 NE or the outflow 
associated with IRS 5. Unfortunately, our submillimeter map is not complete 
in the region surrounding HH 262.

To better show the complex velocity structure within the L1551 outflows, we 
construct position-velocity (p-V) diagrams by averaging the spectra 
perpendicular to the p-V axis over a finite width. In Figure~\ref{fig:pv_irs5} 
we show a p-V diagram 
constructed from the \jone\ data with a width of 2\arcmin.25. The p-V diagram 
is aligned along the main outflow axis (at a position angle of 50\arcdeg) and 
passes through IRS 5 (at zero offset). Figure~\ref{fig:pv_submm} shows the 
p-V diagram constructed from the \jthree\ data along the the same axis with a
width of 1\arcmin.1.

In the red lobe of the main flow, the velocity structure is relatively 
simple, with velocity increasing approximately linearly with distance 
from IRS 5 (Figure~\ref{fig:pv_irs5}).  Such velocity
structure is often dubbed as "Hubble" flow and corresponding feature
in the p-V diagram a Hubble wedge. The velocity field within the
Hubble wedge shows a systematic acceleration, with both the terminal
flow velocity as well as the mean velocity both increasing within the
wedge.  The terminal velocity in the redshifted lobe is reached near
HH 262 at an offset of approximately +11\arcmin\ from IRS 5, where
outflow terminates at all velocities, although cloud emission
continues.  

The velocity field in the blue lobe of the main flow is more structured than 
in the red lobe. The inner region of the outflow is shown in the p-V plot 
of the \jthree\ transition of CO, Figure~\ref{fig:pv_submm}. In this higher
resolution p-V diagram, IRS 5 is clearly the symmetry point between
the redshifted and blueshifted outflow lobes. In the blue lobe,
Hubble wedges associated with HH 29 and HH 259 can be distinguished.
The more extended, but lower resolution p-V diagram, constructed from 
CO \jone\ transition, shows possible additional Hubble wedge features near
HH 29, HH 259, HH 28 and HH 256. The presence of multiple Hubble
wedges has been interpreted as due to either episodic or multiple
outflow events \citep{ag_episodic}.  Low velocity outflow emission is
detected out to an offset position of -18\arcmin, near the edge of the
molecular cloud. Thus, the blue flow may be escaping the cloud, while
the red flow is stopped within the molecular cloud. The different
velocity structure of the red and blue lobes may reflect the differences
in the underlying ambient gas distribution.

We also made p-V plots (not shown here) along the L1551 NE jet axis, and these 
plots are very similar to those shown in Figures~\ref{fig:pv_irs5} and 
\ref{fig:pv_submm}. However, in these plots the location of L1551 NE is 
clearly offset relatively to the symmetry point of the inner redshifted and 
blueshifted Hubble wedges.

\subsubsection{HL Tau Flow}

The most prominent feature associated with the HL Tau flow is the
clam-shell morphology of the blueshifted CO emission seen in
Figure~\ref{fig:mosaic_blue}a. This feature is centered on the HL Tau
jet axis (PA $\sim$ 50\arcdeg, \citet{mundt90}) and opens toward HH
266.  It is likely that this feature reflects the collective impact of
several flows on surrounding gas.  As mentioned earlier, it is
problematic to define the redshifted emission associated with the HL
Tau flow since it overlaps main flow's redshifted emission. 
Figure~\ref{fig:mosaic_red}b shows a linear feature
extending north-east from the HL Tau region and well aligned with the
direction of the HH 30 jet (PA $\sim$ 31\arcdeg, \citet{mundt90}).
The resolution of our \jone\ data does not allow us to study the
details of the HL Tau flow and the \jthree\ map did not extend to this
region.

\subsubsection{EW flow}

In Figure~\ref{fig:EW} we show a mosaic of integrated intensity maps
within successive 4~\kms\ wide redshifted velocity intervals for the
EW flow.  We have not overlaid the \Ha\ optical image on the EW flow,
because the EW flow extends well beyond the coverage of the optical
image.  The portion of the EW outflow covered by the optical image,
with exception of HH 102, shows no bright optical features associated
with the flow. The EW flow has a length of approximately 21\arcmin\
(corresponding to 0.85 pc at 140 pc distance) and has a relatively
large velocity extent ranging from about 8~\kms\ to 20~\kms.  This
flow is highly collimated at all velocities.  The EW flow appears to
originate near HH 102 and at successive higher velocities the emission
is further offset from the origin.  The Hubble flow character to the
velocity structure of EW flow is also shown in the p-V plot
constructed from the \jone\ CO data, Figure~\ref{fig:pv_ew}.  The p-V
plot passes through HH 102 (the zero offset position) and is at
position angle 270\arcdeg.  The outflow emission in the EW flow is
much weaker than in the main flow, so the p-V plot is much noisier.
However, there is evidence for a Hubble flow feature that terminates
at a V$_{LSR}$ of $+16$~\kms\ at an offset position of approximately
16\arcmin.

We speculate that the blueshifted counterpart to the EW red flow is
the feature seen in Figure~\ref{fig:mosaic_red}a extending into the
red lobe of the main outflow. If this feature is related to the EW
flow, the redshifted and blueshifted flows are not co-linear and
would require that the blueshifted portion of the flow curves
northward. The corresponding blueshifted feature that we tentatively
identified with EW outflow is not seen in the p-V diagram since it curves 
north and does not lie along the east-west axis. 
Figure~\ref{fig:l1551_first}, however, provides a good overview of the
possible connection of these two velocity features.

It is intriguing that the high velocity dense gas clumps detected in
CS by \citet{ps95} are located near L1551 NE and HH 102.  Based on the
interferometer results, \citet{ps95} and \citet{yokogawa03} suggested
that NE was formed in a swept-up shell of gas produced by the IRS 5
outflow.  The spatial and velocity structure of the CS emission
toward HH 102 is similar to that near NE.  We speculate, that like
NE, the outflow from IRS 5 may have triggered another star formation
event within the dense shell of swept-up gas near HH 102. Additional evidence 
for the interaction of the IRS 5 jet with material in this region is provided 
by the optical spectroscopy of HH 264, presented by \citet{hartigan00}. In 
fact, recently, \citet{moriarty2006} detected relatively strong
far-infrared dust emission at 850 $\mu$m toward HH 102. Due to 
insufficient spatial resolution they were not able to conclude if there 
is any point source within the extended structure. Thus, a
newly formed star in this region may be responsible for producing the
EW flow and its blueshifted counterpart.

\section{Mass and Energetics}

The L1551 cloud contains over a dozen young stellar objects
concentrated in a region less than 1 pc in diameter and many have
associated HH objects or molecular outflow activity, all of which
attest to the overall complexity of the L1551 region. We define four
areas of interest to study the molecular outflow activity and these
are shown in Figure~\ref{fig:mv_regions}.  Since the mass distribution
along the outflow can be a good indicator of the entrainment
mechanism, the idea is to isolate the various outflow components as
much as possible. Regions A1 and A2 mark the blue and redshifted
lobes of the main outflow.  We also separately analyze the
blueshifted emission in region A2, which may be associated with the
EW outflow.  Region A3 delineates the redshifted and blueshifted
emission likely associated with HL Tau outflow and region A4 the
redshifted emission associated with the EW outflow.

\subsection{Excitation Temperature}

The excitation temperature (T$_{\rm ex}$) of the high velocity gas can be
estimated from the ratio of \twelveCO\ \jthree\ to \twelveCO\
\jone\ line, both available over a significant part of
areas A1 and A2.  The \twelveCO\ \jthree\ data have been spatially
convolved to match the spatial resolution of the \twelveCO\ \jone\
data, and \twelveCO\ \jone\ data have been spectrally smoothed to
match the 0.9~\kms\ velocity resolution of the \twelveCO\ \jthree\
data.  Using positions where both \jone\ and \jthree\ data exist and
where there is a significant outflow emission, we compute an average
spectrum for regions A1 and A2; these spectra and their ratio are shown in
Figure~\ref{fig:texc_fit}. At the velocity of the
ambient cloud emission the ratio, R = T(\jthree)/T(\jone), has a minimum
value close to the ratio of R$=0.58$ predicted for optically thick,
thermalized gas at temperature of 10~K.

With the increasing velocity, ratio R increases and becomes relatively 
constant at outflow velocities.  The weighted mean ratio in region A1 from 
V$_{LSR}$ of $2$ to $-9$~\kms\ (blueshifted) is R$ = 1.69\pm0.16$, and in 
region A1 from V$_{LSR}$ $10$ to $23$~\kms\ (redshifted) is R$ = 1.71\pm0.15$.
  If we assume that the line wing emission is optically thin, and ignore 
the CMB background term in the radiation equation, the line ratio R between 
any two transitions, can be related to the  T$_{\rm ex}$ as follows:

\begin{equation}
R = \frac {\nu_{\rm J_2}^2}{\nu_{\rm J_1}^2} exp(- \frac {h}{2kT_{\rm ex}} [\nu_{\rm J_2}(J_2 +1) - \nu_{\rm J_1}(J_1 +1)])  
\end{equation}
where $J_2$ and $J_1$ are the upper state quantum numbers of the two 
transitions, respectively; $\nu_{\rm J_2}$ and $\nu_{\rm J_1}$ are the 
frequencies for the two transitions. In the case of the ratio R of the 
\jthree\ and \jone\ transitions, we derive: 

\begin{equation}
T_{\rm ex} = \frac {-27.7}{ln(R/9)}  
\end{equation}

If the emission were optically thin, then the excitation temperature
for both regions A1 and A2 would be approximately $T_{ex}$ = 16.5 K.
However, if the gas is not optically thin, then this ratio provides
only a lower bound to $T_{ex}$.  For example, if the optical depth of
the \jthree\ line of CO is of order 2, then the observed ratio would
imply an excitation temperature of $T_{ex}$ = 25 K instead of 16.5 K.
In the next section we will show that over the most of the outflow 
velocity extent, the CO \jone\ line has moderate optical depths ($\tau
= $ 1--2), larger in the near line wings and slowly decreasing to
higher outflow velocities. It is surprising that the ratio R is
relatively unchanged as a function of velocity since the optical depth
of the gas decreases at higher outflow velocities.  Therefore if there
is any change of T$_{ex}$ with velocity, it must decrease from lower
to higher outflow velocities.  Although the ratio R provides only a
lower limit to the excitation temperature, the optical depth is
relatively small, so the excitation temperature is unlikely to be
grossly underestimated.  

\subsection{Outflow Mass and Energy}

Ideally, the mass of outflowing molecular gas should be determined
using an optically thin line such as \thirteenCO\ or C$^{18}$O.
However the emission in the rarer isotopic lines of CO is weak, and
thus the spatial and velocity extent to which the outflow could be
traced would be extremely limited.  In early studies of outflows, the
mass was often derived from \twelveCO\ without corrections for optical
depth.  Even after it was recognized that the \twelveCO\ emission in
the outflowing gas was not optically thin, a single correction for
optical depth was used at all velocities.  More recently,
\citet{brlb99} and \citet{ybb99} introduced a velocity dependent
opacity-corrected column density approach, that has been used with 
modifications in many recent observational studies such as
\citet{ag_mvfit}.  This type of correction is essential for evaluating
the outflow mass as a function of velocity.

Since an individual \thirteenCO\ spectrum does not have the signal to
noise to detect the outflowing gas, we need to average over large
regions of the outflow.  This is the same procedure as employed by
\citet{brlb99} and \citet{ybb99}.  In each of the regions previously
defined (see Figure~\ref{fig:mv_regions}), we average over the extent
of the outflow emission and form an average \twelveCO\ and
\thirteenCO\ spectrum.  For areas A1 and A2 these average spectra
are shown in Figure~\ref{fig:averaged}.  At velocities where both
\twelveCO\ and \thirteenCO\ are detected with at least 3$\sigma$
certainty, we form the ratio (Figure~\ref{fig:averaged}). Similar 
results are obtained for A3 and A4 regions.

In \citet{brlb99}, \citet{ybb99} and \citet{ag_mvfit}, they fit the 
ratio points with a
parabolic function in order to extrapolate the optical depth beyond
the velocity extent of detectable \thirteenCO\ emission.  Examination
of the ratio trends for regions A1 and A2 show that they would not be
well fit by a parabolic function.  The ratio remains flat in the high
velocity wings, implying that the moderate optical depths persist to
relatively large velocities.  The largest measured ratios are $\sim
30$, considerable smaller than the expected value of isotopic ratio 
of approximately 62 for optically thin lines \citep{lp93}.  Instead of 
a parabolic function, we have
fit the ratio with a logarithmic function of the form:
R$_{12/13}(\vel) = Aln|B-\vel| + C$, where $A$, $B$ and $C$ are fitted
parameters, and $\vel$ is the velocity.  This functional form fits
reasonably well the variations in the observed ratio from the line core to 
the line wing velocities, Figure~\ref{fig:averaged}. We use the 
logarithmic fits to correct for the optical depth in \twelveCO\ line.

We derive the column density of the gas in the outflow by a two step process.
Assuming that \twelveCO\ emission is optically thin we use the following 
equation, at each position:

\begin{eqnarray}
N^{thin}_{12}(\vel)=1.15 \times 10^{14}\frac{(0.36T_{\rm ex}+0.33)\int T^{12}d\vel}{e^{-T_0/T_{\rm ex}}(1-0.15(e^{T_0/T_{\rm ex}}-1))}
\end{eqnarray}  
where T$_0$ is the value for h$\nu$/k, equal to 5.53~K for the
\twelveCO\ \jone\ line, and the velocity is in \kms.  The derived
optically thin column densities can then be corrected for optical
depth effects using the following expression:
\begin{eqnarray}
N^{thick}_{12}(\vel)=N^{thin}_{12}(\vel)\frac{62}{R_{12/13}(\vel)},
\end{eqnarray}
where R$_{12/13}$ is determine from the logarithmic fits as a function
of velocity, and we assume an isotopic abundance ratio of 62.

The outflow mass as a function of velocity in each pixel is then computed from
$M(\vel)=2\mu m_{H}A N_{H_2}(\vel)$, where $\mu=1.36$ is the mean
hydrogen mass accounting for He and other molecular constituents,
m$_{H}$ is the mass of the hydrogen atom and A is the physical area of
one pixel at the distance of the source. N$_{H_2}$ is the molecular
hydrogen column density obtained using the relation
N$_{H_2}=1.1\times10^4$N$_{12}$ by \citet{flw82} for the Taurus cloud; this 
result is consistent with more recent determinations, summarized by 
\citet{hlh04}, for other nearby dark clouds.
Table~\ref{tbl:mass} summarizes our results for the mass of the
outflowing gas in the various regions previously defined, tabulated
for both $T_{\rm ex}=16.5$~K and $T_{\rm ex}=25$~K.  Gas at
velocities $\leq 5$~\kms\ and $\geq 8$~\kms\ is considered blueshifted
and redshifted respectively. The total outflow mass in each area A1-A4,
is obtained by summing up calculated mass over all outflow velocities.
The total outflow mass in blueshifted
and redshifted emission are denoted as BT and RT in the Table~\ref{tbl:mass}. 
The
redshifted emission in A2, A3 and A4 do not overlap, so RT is simply
the sum of all redshifted mass, momentum and energy of A2~R, A3~R and
A4~R. However, there is some overlap of blueshifted emission in A1~B
and A2~B (see Figure~\ref{fig:mv_regions}), which we have accounted
for in our mass estimate labeled BT.

In Table~\ref{tbl:mass}, we also give estimates for total momentum P =
MV$_{out}$, and kinetic energy, E$_K$ = 0.5MV$_{out}^2$ in the
molecular outflows.  We have made no attempt to correct for the
inclination of the outflow, so these are only lower limits to the true
values of momentum and kinetic energy.  Not correcting for the
inclination of the outflow, will not affect the mass estimate, since
accounting for the tangential velocity will only rescale the velocity
axis. The highest uncertainty in the mass estimates, factor of 2,
comes from the uncertainty in the N$_{H_2}$ to N$_{12}$ ratio
\citep{flw82}. In addition, the outflow mass at the lowest velocities,
may have a contribution by the ambient cloud mass.  The ambient cloud
contamination will decrease with the outflowing velocity. We consider
the outflow to start at $\sim$1.5~\kms\ from the host cloud's mean
velocity both for the blueshifted and redshifted
velocities. Therefore the effect of the cloud contamination to the low
velocity outflow mass is minimal. The choice of the polynomial fit
function, is a result of arbitrary choice, and although logarithmic
function mimics the behavior of \twelveCO\ to \thirteenCO\ emission
ratio better than the parabolic function at the intermediate outflow
velocities, it never reaches the thin limit and if \twelveCO\ emission
becomes optically thin we will overestimate mass at these highest
velocities.

\subsection{Mass-Velocity Power Law Dependence}
\label{sec:mvresults}

It is well established that mass distribution within the molecular
outflow has a power-law dependence on velocity, such that
$M_{CO}(\vel) \propto \vel^{-\gamma}$ \citep[see for
e.g.][]{richer2000}.  Observationally the relation is typically
obtained by calculating mass per velocity bin and plotting
log($dM/d\vel$) as a function of velocity offset log($\vel_{out}$)
relative to the host cloud's mean velocity $\vel_0$ ($\vel_{out}=\vel-\vel_0$). 
In a log-log plot, the slope of the
linear fit determines the $\gamma$ index. The value of $\gamma$ is an
important test for proposed mechanisms of molecular outflow
entrainment.

The mass spectra and derived power law indices for regions A1
through A4 are shown in Figure~\ref{fig:mv}. We bin the data in
velocity bins uniformly spaced in log scale. The velocity width for each
point varies from 0.25~\kms\ on the left side of the plot to $\sim
5$~\kms\ on the right side. This has the beneficial effect of
increasing the signal-to-noise at high velocities, where the mass is
decreasing. Propagation of error is carried out as mass points
are binned together, and the error bars for each point are also
plotted in the figure. The error bars represent the statistical
uncertainty in the mass estimates that comes primarily from the rms
noise in each measured spectra and uncertainty in the logarithmic fit 
parameters. There are also systematic errors such as the
uncertainties in the excitation temperature, the functional form of
the logarithmic fits to optical depths (Figure~\ref{fig:averaged}), 
and the uncertainty in the N$_{H_2}$ to
N$_{13}$ (this latter uncertainty does not affect the slope of the
\mv\ law). For low velocities, the errors are small enough that the
error bars are not visible in Figure~\ref{fig:mv}. Because the
statistical errors are very small, but systematic errors could be
large, we chose to fit the data to a straight line, assuming uniform 
weights for all points (Figure~\ref{fig:mv}). If we used the 
statistical errors in the fit, the uncertainties at low velocities 
are much lower than at higher velocities, resulting in $\gamma$ values 
with formally low uncertainties (typically
$\lesssim 0.05$), but with large chi-square values in the fits.

In regions A1 and A2 (the blueshifted and redshifted lobes of the main
flow), the dependence of \mv\ are best fit with two power-law relations 
(Figure~\ref{fig:mv}). The break in the power law occurs at
$\vel_{out}\sim 8$~\kms\ in A1 one and at $\vel_{out}\sim 10$~\kms\ in
A2.  The redshifted and blueshifted emission in the HL Tau region
(A3) have slightly different \mv\ distributions. While the redshifted
emission has a broken power law, the blueshifted emission extends to
much smaller velocities, and shows a single, steep power-law
dependence. The redshifted lobe of the EW flow (see region A4 in
Figure~\ref{fig:mv}), and its possible blueshifted counterpart (see
the dashed line in region A2 of Figure~\ref{fig:mv}) both show a
single, steep power-law dependence. When there is a broken power-law,
the average slopes for the low-velocity and high-velocity portions of
all outflows in the L1551 region are 1.63 and 2.92 respectively. These
values are significantly lower than $\gamma$ values reported towards
other outflows in several recent papers (see \S\ref{sec:mvdiscuss} for
a more detailed discussion on this point).

\subsection{Cloud Mass and Energy} 

We determine the cloud mass by using the \thirteenCO\ map. The line
center optical depth at each point in the map is derived from the \thirteenCO\
peak temperature. The excitation temperature is obtained at each position 
by solving the radiative transfer equation
for the excitation temperature, assuming \twelveCO\ line to be optically
thick. The column density for the \thirteenCO\ \jone\
transitions is calculated at each map point in the velocity range of 
5 to 8~\kms\ .
We find, the total mass of the cloud using this method to be 110~M$_{\odot}$, 
and is reported in Table~\ref{tbl:mass} in the AC (ambient cloud) row. The 
newly estimated cloud mass is larger than the previous estimate of 
\citet{mss88} and in agreement with \citet{sb80}.
The 
kinetic energy of the cloud is assumed to be the turbulent energy of the cloud,
 and is estimated using E$_{\rm turb} = 3/(16ln2)M_{\rm cloud}\Delta \vel^2 $. 
 Mean turbulent velocity of the ambient gas, $\Delta \vel =$ 1.2~\kms\ is 
determined from several \thirteenCO\ line profiles in the cloud, as the full 
line width at the half maximum. We find that kinetic energy of the L1551 cloud 
is $\sim 8.5 \times10^{44}$~ergs. 

\section{Discussion\label{sec:discuss}}

\subsection{Entrainment Mechanism for CO Outflows in L1551}

The combination of optical H$\alpha$, millimeter and submillimeter CO
data provides a more complete picture of the structure and kinematics
of the molecular outflows in the L1551 region, allowing us to better
investigate the driving mechanism of molecular outflows.  Currently
the two most promising models of entrainment of outflows are the
jet-driven bow-shock model and the wind-driven shell model.
Numerous hydrodynamic and analytical models have treated both types of
entrainment mechanisms and predict distinct morphologies for the
appearance of the integrated intensity maps, position-velocity maps,
and the mass-velocity power law indices \citep[for examples
see][]{ssy97, zz97, lee00, lee01, ostriker01}.

One of the most distinctive observational characteristics of molecular
outflows is the power-law dependence of flow mass with velocity,
$M_{CO}(\vel) \propto \vel^{-\gamma}$. Usually, log(dM/dv) versus log(v) has a
linear dependence, with a single power law index $\gamma$ ranging
from about 0.5 to 3.5, with the majority of observations indicating
$\gamma \sim$2 \citep{richer2000}. At velocities larger than 10~\kms\
from the ambient cloud velocity, many outflows show a break in the
power-law with higher velocities showing a steeper slope. The break 
at higher velocities might indicate that there are two distinct outflow 
components, perhaps corresponding to a recently accelerated component, 
and a slower-moving swept-up component. It
also appears that the low-velocity power-law index of $\sim 2$ is the
same for low-mass as well as high-mass objects, indicating that some
common acceleration mechanism operates over several decades of stellar
luminosity \citep{richer2000}. There is also some indication, especially for
high-mass objects that the slope, $\gamma$ steepens with age
\citep{richer2000}. Thus, observations of $\gamma$ and the occurrence of
breaks in the power law could thus be used to constrain the driving
mechanism in a given outflow. In turn, the trends seen in the mass versus 
velocity (\mv) power
laws of observed outflows can be used to refine the theoretical models
of outflow entrainment. Accurate determination of the value of
$\gamma$ is thus quite important.

Numerical simulation of both jet- and wind-driven models \citep{lee01} 
show that the power law index in \mv\ plots for jet-driven bow-shock systems 
ranges from 1.5 to 3.5, while the wind models yield smaller value of $\gamma$ 
in the range of 1.3 to 1.8. In the jet-driven models, $\gamma$ is a strong 
function of inclination, with lower values of $\gamma$ obtained when the 
outflow has larger angles of inclination to the plane of the sky. 
\citet{zz97} show that power-law with break is produced for the 
jet-driven entrained gas. 

Jet-driven bow-shock models show Hubble-law like features in the
position-velocity plots with the highest velocities at the ''hot
spots'' (head of the jet) and decreasing velocity trend toward the
source (wings of the bow shock). The p-V structure is associated with
the broad range of velocity near the bow tip while there is a small
and almost constant velocity in the bow wings, often producing a
convex spur structure along the jet axis at the highest velocities
\citep{lee00, lee01}. Figure 20 in \citet{lee00} shows a strong dependence 
on inclination angle for the p-V diagram structure in bow shock model. In
the case where several bow shocks are driven because of the presence
of a pulsed jet, several spur structures can be seen in the p-V
diagram. \citep[see Fig. 7,][HH 212]{lee00}. On the other hand,
wind-driven entrainment mechanism do not show any evidence for spur
features in the p-V plots; the p-V plots are typically parabolic in
shape with the vertex of the parabola coincident with the location of
the source. The morphology of the shape is somewhat inclination
dependent, but the p-V plots are quite distinctive in appearance from
that produced by jet-driven models.

In \S~\ref{sec:mvresults}, we report our results for the mass-velocity
power laws. Our main result is that when there is a break in the power
law, the power law slope for low velocities is shallow, $\gamma\lesssim
2$. The power law index for high velocities is $\gtrsim 3$ (see
Figure~\ref{fig:mv}).

The main CO outflow oriented at 50\arcdeg\ from IRS~5 shows mostly
jet-driven bow-shock features, but also shows some morphological 
features seen in wind-driven flows. The \mv\ plots for the blue and
redshifted lobes (regions A1 and A2) of the main IRS 5 outflow show a
power-law break as predicted by a jet-driven flow (see
Figure~\ref{fig:mv}). The millimeter and submillimeter p-V plots for
IRS 5 (Figures \ref{fig:pv_irs5} and \ref{fig:pv_submm}) show multiple
Hubble-law wedges, with spur-like features at the highest velocities
(see especially the submillimeter data in Figure~\ref{fig:pv_submm}).
The blueshifted side of the CO \jone\  p-V plot (Figure~\ref{fig:pv_irs5})
might be interpreted as a parabolic feature, or it may be two Hubble
like wedges adjacent to each other. But the redshifted side shows no
evidence for any parabolic structure. A single convex spur-like
feature is seen in the p-V plot on the redshifted side. The channel
maps in Figures \ref{fig:mosaic_blue} through
\ref{fig:mosaic_red_submm} show a morphology where the highest
velocity emission is the most collimated, and is along the projected
jet axis, and is often seen in the vicinity of HH objects, probably
highlighting the interaction regions of shocks at the heads of the
bow-shock interfaces of the underlying jet with the swept-up molecular
material. The overall spatial morphology of the low-velocity gas in
blue and redshifted lobes is roughly parabolic (see Figures
\ref{fig:mosaic_blue} through \ref{fig:mosaic_red_submm}), with the
slower swept-up gas manifesting itself as limb-brightened shell-like
features. It has been proposed that the entrainment mechanism itself
may evolve in outflows, with the early stages being a collimated
jet-driven flow, which in later stages evolves into very wide-angle
wind \citep{velusamy98, richer2000}. If so, the characteristics of
both jet-like and wind-driven entrainment for the IRS 5 flow might be
indicative of its relatively evolved stage as an outflow.

The EW flow on the other hand shows only jet-like features. The flow
is highly collimated (see Figure~\ref{fig:EW}), the \mv\ plots for both the
redshifted western component and the possible blueshifted 
component (in the redshifted lobe of IRS~5 flow) show single sloped
power law with steep power-law indices ($\gamma=3.1$ and $4.5$
respectively) indicative of jet-driven bow-shock entrainment. The p-V
diagram for the EW flow (Figure~\ref{fig:pv_ew}) shows no parabolic
features, but a low-intensity Hubble-like wedge ending in a broad spur
at the tip. It is noteworthy that the \mv\ plot (see
Figure~\ref{fig:mv}) for the EW flow is quite distinct from that of
the IRS 5 flow. As suggested earlier, it is possible that the EW flow
is dynamically younger flow, driven from an hitherto undetected
source close to the location of HH 102 at about the location of the
850~$\mu$m peak emission seen by \citet{moriarty2006}. Unlike the IRS 5
flow, no optical counterpart HH objects have yet been detected along
the EW jet axis.

\subsection{Systematic Effects on Mass Spectra Slopes\label{sec:mvdiscuss}}

The power law indices at both low and high
velocities that we derive for L1551 are not as steep as some recent
determinations of $\gamma$ in other outflows. For example, other
groups have found $\gamma > 3.5$ (single power-law) for Circinus
\citep{brlb99} and Barnard 5 \citep{ybb99}, broken power-law with
$\gamma \sim 2$ (low velocities) and $\gamma \gtrsim 5$ (high
velocities) in OMC-2/3 \citep{yu2000}, broken-power law with
$\gamma\gtrsim 3$ (low velocities) and $\gamma > 6$ (high velocities)
for HH 300 \citep{ag_mvfit}, broken power-law with $\gamma \sim 2.2$
(low velocities) and $\gamma \gtrsim 3.5$ (high velocities) for PV Cep
\citep{ag2002}.  These recent studies have reported systematically
steeper power law indices than reported before, and when accompanied
with a break in the power law, much steeper slopes at higher
velocities.  One common feature of these recent results with larger
values of $\gamma$ was that they employed a velocity-dependent opacity
correction. In this work, we have employed a similar
velocity-dependent opacity correction approach, but we derive smaller
values of $\gamma$. Our results here seem to counter this trend
towards larger values of $\gamma$.  Does this indicate that L1551 is a
different kind of outflow system than these other outflows?  There are
several systematic effects in the derivation of mass-velocity spectra
that if not properly accounted for, can result in errors in the
determination of $\gamma$. Below we consider these systematic effects
to explain the difference in $\gamma$ derived in this paper compared
to some recent determinations of this quantity.
 
(1) The overall mass estimate at all velocities can be affected by the
size of the outflow mapped. If high velocity portions of the outflow
are missed in the mapping, the derived $\gamma$ would be an
over-estimate. Our study as well as the studies of \citet{brlb99,
  ybb99, yu2000, ag_mvfit,ag2002} have mapped a larger area of the
outflow lobes than most previous studies (primarily due to the higher
sensitivities of receivers and the advent of OTF mapping). The size of
the mapped region is probably not the cause of difference in derived
value of $\gamma$.

(2) The exact form for the functional fit to the velocity dependent
opacity correction can affect the slopes of the mass spectra. In the
studies of \citet{brlb99, ybb99, yu2000, ag_mvfit,ag2002}, parabolic
fits were used to extrapolate the value of the ratio
($^{12}$CO/$^{13}$CO) beyond values where the ratio is well
determined. We use a logarithmic function (see
Figure~\ref{fig:averaged}) instead. The logarithmic function is a
better fit to the variation of the ratio seen in our data, but it
tends to be a more slowly varying function with velocity than the
parabolic fits. This makes for larger corrections (more mass) for the
higher velocity points, which in turn makes the mass spectral slopes
shallower. It should be noted that, compared to earlier studies, our
data have higher sensitivity even at higher velocities from the
systemic velocity, and this leads to a better constraint of the
$^{12}$CO/$^{13}$CO ratio to much higher velocities. The variation of
optical depth of CO in molecular outflows with velocity has not been
well determined to high velocities. However, when high signal-to-noise
ratio spectral data have been obtained towards high velocity molecular
outflows, the ratio of $^{12}$CO/$^{13}$CO seems to be slowly varying
with velocity \citep{snell84}, as is seen in our study.  As another
example, in Figures 7(c) and 7(d) of \citet{ag2002}, it is seen that
the ratio of $^{12}$CO/$^{13}$CO seems to be actually flat, however,
the authors use a parabolic fit. A parabolic fit is a much faster
extrapolation function in velocity and predicts that the gas gets to
the optically thin limit at lower velocity offsets from the systemic
velocity.

(3) Previous studies of mass spectra apply a method of only using data 
which exceeds a specified threshold in rms value of $T_A^*$. For example,
\citet{ag_mvfit} used a 2$\sigma$ threshold, while
\citet{brlb99} used a 1$\sigma$
threshold. Figure~\ref{fig:rms_vs_norms} shows the effect of using a
rms threshold in the determination of the mass spectra. The rms
thresholding method preferentially rejects points at higher velocity
channels, since those points have the lowest signal to noise, and
hence do not meet the rms threshold. Because of this, mass is
under-estimated preferentially at higher velocities, thereby
artificially steepening the mass-spectral slopes.

The velocity-dependent optical depth correction used in \citet{brlb99,
ybb99, yu2000, ag_mvfit,ag2002} as also in this work does result in
more accurate mass determinations. In particular, not using the
correction results in an under-estimation of the mass at especially
lower velocities, artificially giving low mass-spectra slopes.  On the
other hand, care must be taken with the functional form of the fit of
optical depth with velocity. Figure~\ref{fig:rms_vs_norms} shows the
importance of using all the data or at least being aware of the effect 
that posing a specified threshold in rms value will have on mass spectrum.

\subsection{Combined Effect of the Outflows on the Cloud}

The mass for the L1551 cloud estimated from the fully sample \thirteenCO\ map
at line core velocities of 5 to 8~\kms\ is $\sim 110$~M$_{\odot}$. Outside of 
the  line core velocities, the total mass estimated in the outflow is 
$\sim 5.3$~M$_{\odot}$ for $T_{\rm ex}=16.5$~K or $\sim 6.7$~M$_{\odot}$ for 
$T_{\rm ex}=25$~K. Therefore the mass in the outflow is approximately 5-6.5$\%$ 
of the mass of surrounding cloud, depending on the adopted value for the 
$T_{\rm ex}$. The total momentum and energy carried by the outflow, estimated 
from only the line of sight (radial) velocity, is 20~$M_{\odot}$\kms and 
1.5$\times10^{45}$ ergs, respectively. It it usually assumed that the L1551 
outflows have small inclination angles relatively to the plane of 
the sky, so there is a large inclination correction to the momentum and 
energy determinations. However, since the inclination angle is extremely 
uncertain, we will use the values quoted above as lower limits to the true 
values.

Many authors suggest that outflows provide a mechanism to regulate star 
formation
by removal of the gas from star-forming regions, especially in the regions 
that lack massive stars (eg. \citet{mm00}, and references therein).   
For the outflow entrained gas, to escape from the cloud, it must be moving at 
least at the escape velocity. Using a mass of 110 M$_{\odot}$ and $\sim
1$~pc radius for the cloud, the escape velocity is calculated to be
$\sim 1$~\kms. If all of the momentum of the outflowing molecular gas was 
transferred to the cloud uniformly, then the gas would not escape. However, 
since outflows are highly collimated, momentum is transferred non-uniformly 
and much of the present outflowing molecular gas is likely to escape, unless
it is slowed down by encounters with ambient gas. Our observations suggest 
that in fact the main blue lobe has broken out of the cloud near HH 256, 
and that their is insufficient column density of gas to slow the gas as 
it escapes. However, in the main red lobe, the molecular outflow is still well
within the boundaries of the cloud and it remains to be determined if
there is sufficient column density of gas to halt the outflow.
Thus, the outflows are unlikely to disperse the entire cloud, but in local 
regions will be a major agent for disruption.

The total binding energy of the cloud is given by 
$\sim GM^2_{\rm cloud}/R $. Using a cloud mass of 110 M$_{\odot}$ and 
$\sim 0.5$~pc for the cloud radius, we find the binding energy to be 
2$\times 10^{45}$~ergs. Thus, the energy of the outflow 
1.5-2$\times 10^{45}$~ergs, is comparable to the binding energy of the cloud,
even if no correction for the inclination angle is made. 
It has been suggested that parsec-scale outflows can be a source of energy 
to replace the energy dissipated in turbulence and thus provide the power
to maintain the magnetohydrodynamic (MHD) turbulence \citep{ln06}. 
The rate of turbulent energy dissipation can be estimated from equation (7) 
in \citet{maclow99}.  Assuming a cloud mass of 110~M$_{\odot}$ and a mean
turbulent velocity dispersion of 0.45~\kms\ (estimated from the \thirteenCO\ 
line width), we find that the energy dissipation rate of turbulence in L1551 
cloud is 0.003~L$_\odot$. Thus, if even a small fraction of the outflow energy 
could be coupled 
into cloud turbulence, the outflow could sustain the turbulence for over a 
million years, much greater than the lifetimes of molecular outflows.

We have shown that the molecular outflows contain substantial kinetic energy.
However, the effect outflows have on their parent clouds depends on the 
efficiency in which momentum and energy can be transferred from the outflowing 
molecular gas
to the surrounding cloud material.  Currently, the observations are
insufficient to  answer this question, and there are no good theoretical 
models that address the effect outflows have on disruption of the cloud or 
on their role in replenishing the reservoir of turbulence energy in clouds.  
However, our work and several recent studies of parsec scale outflows,
 suggest that young stars, through the combined effect of their outflows, 
may have an important impact on the dispersal of their parent molecular cloud 
and to power turbulent motions.

\subsection{Sequential Star Formation in L1551?}

The L1551 molecular core is an active star-forming region which
harbors at least 2 small clusters of protostars, one centered in the
IRS~5 and L1551 NE region, and another in the HL~Tau region. These two
mini-clusters are clearly undergoing active star-formation
characterized by HH objects, optical and near-infrared jets. In
addition, multi-wavelength surveys in mid infrared \citep{galfalk04},
X-ray \citep{favata03}, optical and near-IR \citep{briceno98}, optical
spectra \citep{gomez92}, and in H$\alpha$ \citep{gncg92} have shown
that the whole L1551 molecular core is surrounded by a halo of young
stars in the Class II stage or older, indicating that star-formation
activity has been ongoing in this region for at least a few million
years.  Here, we explore the possibility that the current epoch of
star-formation activity in L1551 may be the result of sequential
star-formation induced by the effects of the multiple outflows in the
system.

The protobinary system in IRS 5 has two active jets. The current
direction of either of these jets is not aligned with the
$\sim50$\arcdeg\ oriented main CO outflow in this region, the latter
being driven by the source(s) at IRS 5. At least at some point in the
past, the predominant direction of the jets from the IRS~5 system was
probably aligned with the bulk molecular outflow direction in
L1551. 
It has been suggested that L1551 NE was formed in the swept-up
shell of gas produced by the IRS 5 outflow \citep{ps95,
yokogawa03}. L1551 NE is itself a multiple stellar system, and
possesses optical jets, and is thought to power a large-scale HH flow
including HH 454, HH 28 and HH 29 \citep{drb99, hartigan00}. The
evidence for a CO outflow from L1551 NE is not as clear-cut despite
many observational attempts to delineate the outflow, mostly because
of its similar alignment and close proximity to the main flow from IRS
5. The high angular resolution interferometric observations of
\citet{ps95} and \citet{yokogawa03} show two symmetric CS structures,
one eastward from IRS~5 near L1551 NE, and thought to be responsible
for the formation of L1551 NE, and another near HH 102. As discussed
before, with the presence of a 850 $\mu$m dust-continuum clump near HH
102 \citep{moriarty2006}, it is tempting to suggest that a similar
triggering mechanism can be attributed to the EW outflow. It is
possible that the same bipolar jet streaming eastward started the
star-formation process in L1551 NE, and streaming westward, started
the star-formation process in the clump near HH 102. This new driving
source near HH 102 is maybe responsible for the EW flow. This new
protostar is obscured in the HH 102 region, and the highly collimated
CO outflow, and the jet-like characteristics of the EW flow are indeed
consistent with a relatively young protostellar system. Follow-up high
angular resolution observations of the HH 102 region would allow us to
probe the possibility that there is a new embedded source near HH 102.

The evidence for sequential star-formation activity is further
strengthened by some recent observations of the L1551 region. A new
pre-protostellar clump, dubbed L1551-MC has been discovered near HH
265 \citet{swift05} using NH$_3$ observations, and later confirmed
\citep{moriarty2006} by 850 $\mu$m continuum observations. L1551-MC is
located at the edge of HH 265, which may be a HH object at the end of
a jet arising from a driving source within the HL Tau complex (see
Figure 17 of \citet{moriarty2006}). It is possible that the next
generation of star-formation in the L1551 molecular cloud could occur
at the L1551-MC region.

 \section{Summary And Conclusions}

All L1551 molecular outflows have been mapped at high sensitivities in
the \twelveCO\ and \thirteenCO\ \jone\ transitions. Follow-up
submillimeter observations in \twelveCO\ \jthree\ emission towards a
good fraction of the L1551 outflow system were carried out as
well. The millimeter and submillimeter data are combined with
large-scale, narrow band, optical H$\alpha$ images in order to perform
a detailed study of the outflows in L1551. The main conclusions of the
paper are summarized below.

\begin{enumerate}
\item The full extent of the main CO outflow is 32\arcmin\ (1.3 pc),
  and is oriented at $\sim 50$\arcdeg\ position angle. The molecular
  outflow extent is similar to the optical extent of the flow seen by
  \citet{drb99}. Outflowing molecular material extends well beyond the
  last known HH object, HH 256 in the south-west, but stops short of
  HH 286 in the north-east. Our molecular line maps indicate that HH
  286 has completely exited the L1551 cloud.
\item Most of the CO molecular outflow appears to be driven by
  source(s) in the IRS 5 region. Some features,
  especially in the redshifted lobe of the outflow may be attributed
  to outflows originating in L1551 NE and the HL Tau region.
\item Our maps cover a much larger extent of the collimated
  redshifted EW flow. The EW flow is $\sim 0.82$ pc in length, and is
  $\sim 0.15$ pc in width. In contrast to earlier studies, the
  sensitivity in our data reveals that the EW flow extends to high
  velocities (6.5~\kms\ to 20~\kms). It is suggested that a smaller
  extent blueshifted component seen in the main northwestern
  redshifted lobe is the blueshifted counterpart of the EW flow, and
  that the driving source for the EW flow is close to the location of
  HH 102.
\item We refine the procedure of velocity-dependent opacity correction
  adopted by recent authors to obtain a more accurate determination of
  outflow mass with velocity. The resultant mass-spectral power law
  indices are lower (less steep) than recently obtained indices
  towards other outflows. We attribute this systematic difference in
  power-law index to the better quality of our data, and the more
  careful approach we use in calculating mass. The resulting mass of
  the L1551 molecular cloud core is $\sim 110$~M$_\odot$ and the
  combined mass of the outflows in L1551 is $\sim 7.2$~M$_\odot$.
\item Even for no inclination angle correction, and lower $T_{\rm ex}$
  assumption, the kinetic energy of the outflow is comparable to the
  binding energy of the cloud and there seems to be enough outflow
  energy to maintain turbulence in the cloud. Therefore
  the parsec scale outflow in L1551 has a potential of making a strong
  impact on the cloud's evolution and fate.
\item Multiple lines of evidence from \mv\ spectral indices, position
  velocity plots, and morphological appearance of our multi-wavelength
  data indicate that the EW molecular outflow is a good
  example of jet-driven bow shock entrainment. The main molecular
  outflow in L1551 shows evidence for entrainment from both jet-driven
  and wind-driven mechanisms.
\item We suggest that new stars are being triggered to form in the
  L1551 system as a result of the effects of the multiple outflows in
  the region.
\end{enumerate}

\acknowledgments We gratefully acknowledge the staff of the HHT for
their excellent support during observations. Dan Logan, Abby Hedden,
Naomi Ridge, and John Howe are thanked for helping with part of the
HHT observations.  This work was supported by NSF grant AST 02-28993
to the Five College Radio Astronomy Observatory.

Facilities: \facility{FCRAO}, \facility{HHT}

\clearpage

\begin{figure}
\caption{Integrated intensity \twelveCO\ \jone\ emission in L1551
overlayed on optical \Ha\ data. Blueshifted gas is integrated in
velocity range of $-20$ to 5~\kms. 
Contours start at 3$\sigma$ and are 1.5, 3, 4.5, 6, 9, 12,
16, 20, 24, 28, 32, 36, 40, 44 K~\kms, plotted in blue. Redshifted
emission is integrated in velocity range of 8 to 20~\kms. Contours 
start at 4.5$\sigma$ level and are
1.5, 3, 5, 9, 12, 15, 18, 22, 26, 30, 34, 38, 42, 46 K~\kms, plotted
in red.  Symbols for different driving sources are introduced: filled
square for L1551 IRS 5, filled circle for L1551 NE, filled triangle
for HL/XZ Tau system and cross for various HH objects. Notation will
be kept on following figures. CO mapping was performed well beyond the
borders of the optical image, reaching a degree squared centered on
IRS 5 source. Dark spots in the optical image are shock-excited gas,
usually associated with various HH objects.\label{fig:l1551_first}}
\end{figure}

\begin{figure}
\caption{Integrated intensity of the \twelveCO\ \jone\ line within
core velocities from 5 to 8~\kms\ and overlayed on \Ha\ data. Contours are
1.5, 3, 4.5, 6, 9, 12, 16, 20, 24, 28, 32, 36, 40, 44, 48, 52, 56, 60
K~\kms, where 1$\sigma$ level corresponds to $\sim$ 0.2 K~\kms. \twelveCO\ 
\jone\ map in line core velocities shows the distribution of the 
molecular gas in the L1551 cloud. The straight
thick lines in the figure outlines the borders of observed CO map.}
\label{fig:12COcore}
\end{figure}

\begin{figure}
\caption{Integrated intensity of the \thirteenCO\ \jone\ emission
overlayed on optical H$\alpha$  data. Gas is integrated in line core
velocity range of 5 to 8~\kms\ and the lowest contour is at 1 K~\kms\
increasing in steps of 1 K~\kms. 1$\sigma$ level corresponds to 0.1 K~\kms.
\thirteenCO\ \jone\ probes the higher
column density regions within the cloud. The edges of the OTF map are
shown here in straight thick lines.}
\label{fig:13CO}
\end{figure}

\begin{figure}
\caption{Integrated intensity \twelveCO\ \jthree\ emission in L1551
shown in dotted line for blueshifted lobe and in solid line for red
shifted emission overlayed on optical \Ha\ data. Blueshifted gas is
integrated in velocity range of $-20$ to 4~\kms\ and the lowest contour
is 5 K~\kms\ ($\sim$10$\sigma$ level) increasing in steps of 7 K~\kms. 
Redshifted emission is
integrated in velocity range of 8 to 20~\kms. Lowest contour is at 5
K~\kms\ (15$\sigma$ level) increasing in steps of 5 K~\kms. Symbols 
for different driving
sources and jets are kept as on Figure~\ref{fig:l1551_first}.}
\label{fig:CO32}
\end{figure}

\begin{figure}
\caption{Channel-maps of the \twelveCO\ \jone\ blueshifted integrated
intensity emission. The CO line is integrated within 4~\kms\ velocity
bin, starting from 5~\kms. The velocity range used for the integration
is given in each panel. The contours start at 2~K~\kms\ (10 $\sigma$) 
and go in
steps of 2~K~\kms\ in panels ``a'' and ``b''. In the lower panels the
step is changed to 1 K~\kms. In each panel, embedded stars are noted
with square (IRS 5 at (0,0) position), circle (L1551 NE) and triangle 
(HL/XZ Tau) and crosses are for HH objects. In the ``d'' panel we assign 
each cross to the corresponding HH object. The same holds for the next 
figure.}
\label{fig:mosaic_blue}
\end{figure}

\begin{figure}
\caption{Channel-maps of the \twelveCO\ \jthree\ blueshifted
integrated intensity emission. The CO line is integrated within 4
~\kms\ velocity bin, starting from 5~\kms. The velocity range used for
the integration is given in each panel. The contours start at 2
K~\kms\ (10 $\sigma$) and go in steps of 5 K~\kms\ in the panels ``a'' 
and ``b'' and in the other two panels step is changed to 2.5 K~\kms.}
\label{fig:mosaic_blue_submm}
\end{figure}

\begin{figure}
\caption{Channel-maps of the \twelveCO\ \jone\ integrated intensity
within 4~\kms\ velocity bins of the redshifted emission in the main
flow's red lobe and the HL Tau region. The velocity range is given in
each panel. The contours in each panel start at 1 K~\kms\ (5 $\sigma$) 
and go in steps of 2 K~\kms\ in panels ``a'' and ``b'' and in other two panels
the step is changed to 1 K~\kms. The (0,0) is at the IRS 5 location. }
\label{fig:mosaic_red}
\end{figure}

\begin{figure}
\caption{Channel-maps of the \twelveCO\ \jthree\ integrated intensity
within 4~\kms\ velocity bins of the redshifted emission in the main
flow's red lobe and the HL Tau region. The velocity range is given in
each panel. The contours in panels ``a'' and ``b'' start at 2 K~\kms\
(10 $\sigma$) and go in steps of 5 K~\kms. In the panels ``c'' and ``d'' the
contour step is 2.5 K~\kms.  The (0,0) is at the IRS 5 position.}
\label{fig:mosaic_red_submm}
\end{figure}

\begin{figure}
\caption{Position-velocity plot for L1551 IRS~5 in CO \jone.  The left
side of the plot shows the p-V plot, made along the
50\arcdeg\ axis going through IRS~5. The width of the cut is 6 pixels
or 2\arcmin.25.  On the right side, is the integrated intensity image
of the L1551 outflow in \jone\ , rotated 50\arcdeg. The dotted
vertical lines outline the width of the p-V swath. The position of the
driving source IRS 5, corresponds to the zero value on the y axis. The
positions of the different HH objects are marked.}
\label{fig:pv_irs5}
\end{figure}

\begin{figure}
\caption{CO \jthree\ p-V plot for L1551 IRS~5. The
  \jthree\ integrated intensity image on the right side is rotated
  50\arcdeg. The width displayed in the integrated intensity image is
  2\arcmin.2, while the width of the p-V cut is 1\arcmin.1 centered
  on IRS 5 (dotted vertical) lines.}
\label{fig:pv_submm}
\end{figure}

\begin{figure}
\caption{\twelveCO\ redshifted integrated intensity emission in the
EW direction within 2~\kms\ velocity bins. The square at (0,0) position marks 
the IRS 5 and the cross marks HH 102 position. The contours in panel 
``a'' start at 0.75 K~\kms\ (6 $\sigma$) and go in steps
of 0.75 K~\kms. In panel ``b'' the step is changed to 0.125 K~\kms, and 
the lowest contour level is 0.5 K~\kms. In the top two panels, the step 
is 0.125 K~\kms and the lowest contour is 0.25 K~\kms.}
\label{fig:EW}
\end{figure}

\begin{figure}
\caption{CO \jone\ p-V plot for EW flow cut along the
 270\arcdeg\ axis emerging from HH 102 (zero on the y axis). The cut is
 90\arcmin\ wide, and on the right side, the integrated intensity
 image made over redshifted velocities, and rotated 270\arcdeg\ is
 shown.}
\label{fig:pv_ew}
\end{figure}

\begin{figure}
\caption{Four different areas A1, A2, A3 and A4 are identified along
the L1551 flow (see text for more).}
\label{fig:mv_regions}
\end{figure}

\begin{figure}
\caption{Upper panels show averaged \twelveCO\ \jone\ (solid line) and
\twelveCO\ \jthree\ (dotted line) spectra for A1 (left panel) and A2
(right panel) regions. \twelveCO\ \jone\ data are smoothed to the 0.9
~\kms\ velocity resolution of \twelveCO\ \jthree\ . In the lower
panels, the ratios of the averaged \twelveCO\ \jthree\ and \twelveCO\
\jone\ spectra are displayed. The weighted mean ratio in each region is
computed using only filled points, i.e. in the line wings where lines are 
less optically thick.}
\label{fig:texc_fit}
\end{figure}

\begin{figure}
\caption{Optical depth profiles. The averaged \twelveCO\ (solid line)
and \thirteenCO\ (dotted line) spectra in A1 and A2 areas are shown in
the upper panels. In region A1, for velocities larger then 2~\kms, we
show spectra in full resolution, and for lower velocities, data are 
binned in 1~\kms\ bins. Similarly, in region A2, for velocities larger then
10~\kms\ data are binned in 1~\kms\ bins. The lower panels show the
logarithmic fit to the ratio in the A1 and A2 regions
respectively.}
\label{fig:averaged}
\end{figure}

\begin{figure}
\caption{The mass-velocity distribution for the outflows within each 
area A1-A4 in L1551. The
\mv\ distribution is plotted as log($dM/d\vel$) vs. log($\vel_{out}$) where
$\vel_{out}$ is outflow velocity $\vel_{out} = \vel-\vel_{LSR}$.  The
mass points are in velocity bins thet are uniformly spaced in 
log velocity (see text for more). For regions A2 and A3, we present data for 
both the blueshifted outflow (triangle points) and the redshifted outflow 
(diamond points) within these regions. The slopes of the linear fits (the 
power law index, $\gamma$) are indicated in the figure.}
\label{fig:mv}
\end{figure}

\begin{figure}
\caption{The \mv\ relation derived for region A2 redshifted gas by using all
  the data without an rms threshold (diamonds), and by setting a threshold 
  of $3\sigma$ for the $^{12}$CO emission (triangles). We show fits to both 
  \mv\ relations with broken power laws.}
\label{fig:rms_vs_norms}
\end{figure}

\clearpage

\begin{deluxetable}{ccccccc}
\tabletypesize{\scriptsize} 
\tablewidth{0pt}
\tablecaption{Outflow Mass, Velocity and
Energy estimates\label{tbl:mass}} \tablewidth{0pt}
\tablehead{ \colhead{T$_{\rm ex}$} & \colhead{25 K} & \colhead{}& \colhead{}& 
\colhead{16.5 K} & \colhead{}& \colhead{}\\
\tableline
\colhead{Lobe} & \colhead{Mass} & \colhead{Momentum} &
\colhead{Energy} & \colhead{Mass} & \colhead{Momentum} &
\colhead{Energy} \\  \colhead{} & \colhead{M$_{\odot}$} &
\colhead{M$_{\odot}$km s$^{-1}$}& \colhead{$\times10^{43}$ (ergs)} & 
\colhead{M$_{\odot}$} & \colhead{M$_{\odot}$km s$^{-1}$} & 
\colhead{$\times10^{43}$ (ergs)}}
\startdata 
A1 B &4.25 & 15.65 & 111.5 & 3.28 & 12.09& 88.33 \\ 
A2 R &1.40 & 7.46 & 60.7 & 1.08 & 5.76 & 46.80 \\
A2 B &0.53 & 1.52 & 6.2 & 0.41 & 1.17 & 4.75 \\ 
A3 R &0.34 & 1.46 & 8.6 & 0.26 & 1.13 & 6.63 \\ 
A3 B &0.23 & 1.06 & 15.6 & 0.18 & 0.82 & 12.02 \\ 
A4 R &0.59 & 1.92 & 9.82 & 0.46 & 1.48 & 7.57 \\ 
\tableline
BT\tablenotemark &4.54 & 15.66 & 116.7& 3.50 & 12.08 &  90.01 \\ 
RT\tablenotemark& 2.33 & 10.84 & 79.12  & 1.80 & 8.37 & 61.00 \\ 
\tableline 
Total & 6.87 & 26.50 & 195.19& 5.30 & 20.45 & 151.01 \\ 
\tableline 
AC\tablenotemark{a} & 110 &  & 85 & & & \\
\enddata 
\tablenotetext{a}{Ambient Cloud. Mass of the ambient cloud is estimated using 
T$_{\rm ex}$ determined in each spatial point from the $^{12}$CO line thick condition. See text for more.}
\end{deluxetable}

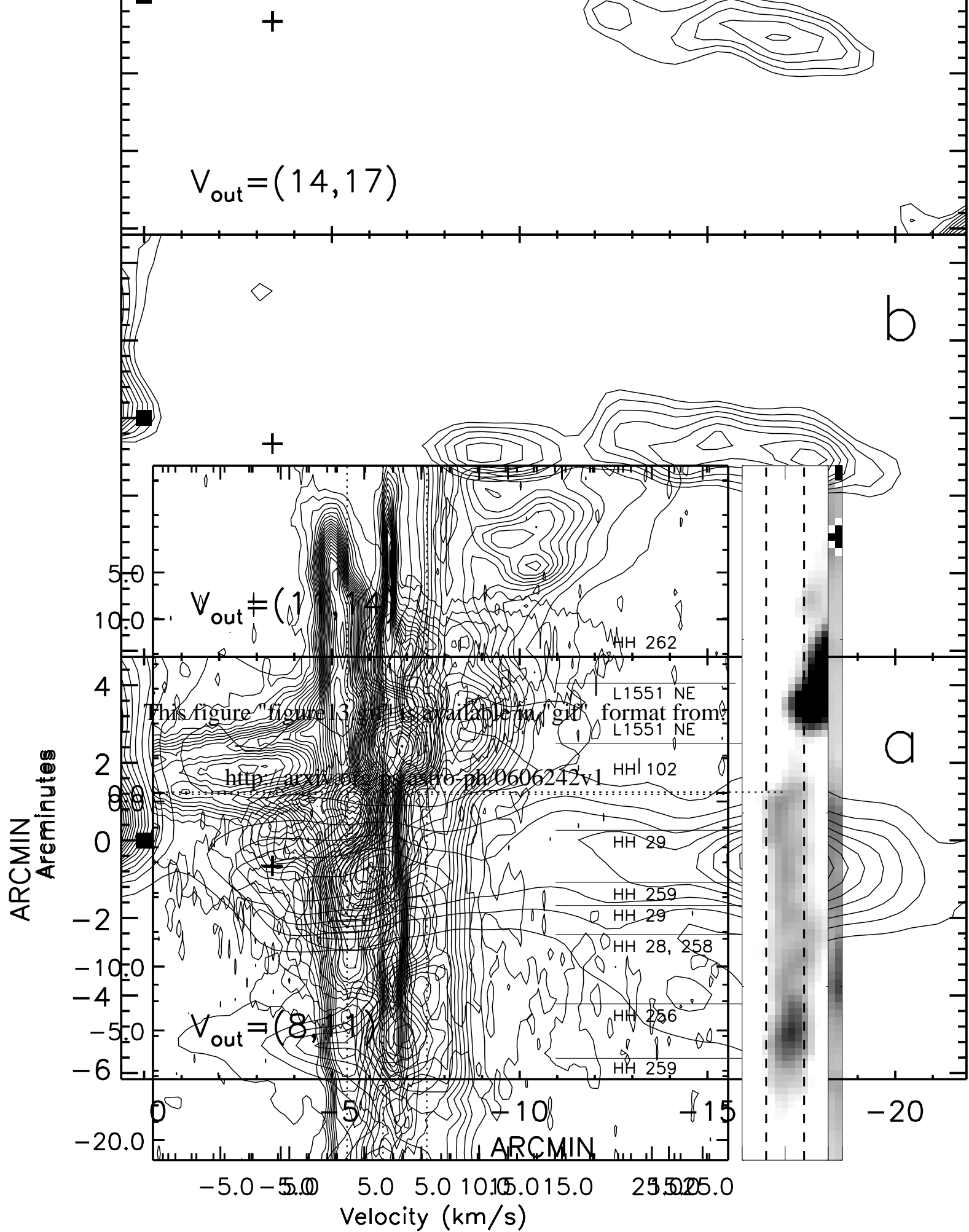
\end{document}